\documentclass[aip,pop,reprint,showpacs,showkeys,amssymb,amsmath,superscriptaddress]{revtex4-1}
\usepackage{graphicx}
\usepackage{subfig}
\usepackage{bm}
\usepackage{color}

\DeclareBoldMathCommand{\bV}{V}
\DeclareBoldMathCommand{\bv}{v}
\DeclareBoldMathCommand{\bx}{x}
\DeclareBoldMathCommand{\by}{y}
\DeclareBoldMathCommand{\bz}{z}
\DeclareBoldMathCommand{\br}{r}
\DeclareBoldMathCommand{\bb}{b}
\DeclareBoldMathCommand{\be}{e}
\DeclareBoldMathCommand{\bB}{B}
\DeclareBoldMathCommand{\bE}{E}
\DeclareBoldMathCommand{\bk}{k}
\DeclareBoldMathCommand{\bA}{A}
\DeclareBoldMathCommand{\bJ}{J}

\newcommand{\V}[1]{\mathbf{#1}}

\begin{document}

\title{Collisionless Reconnection in the Large Guide Field Regime: Gyrokinetic Versus Particle-in-Cell Simulations}

\author{J.~M. TenBarge}
\email{jtenbarg@umd.edu}\affiliation{IREAP, University of Maryland, College Park, MD 20742}
\author{W.~Daughton}\affiliation{Los Alamos National Laboratory, Los Alamos, NM 87545}
\author{H. Karimabadi}\affiliation{SciberQuest, Del Mar, CA 92014}\affiliation{Department of Electrical and Computer Engineering, UCSD, La Jolla, CA 92093}
\author{G.~G. Howes}\affiliation{Department of Physics and Astronomy, University of Iowa, Iowa City, IA 52242}
\author{W. Dorland}\affiliation{IREAP, University of Maryland, College Park, MD 20742}

\date{\today}

\begin{abstract}
Results of the first validation of large guide field, $B_g / \delta B_0 \gg 1$, gyrokinetic simulations of magnetic reconnection at a fusion and solar corona relevant $\beta_i = 0.01$ and solar wind relevant $\beta_i = 1$ are presented, where $\delta B_0$ is the reconnecting field. Particle-in-cell (PIC) simulations scan a wide range of guide magnetic field strength to test for convergence to the gyrokinetic limit. The gyrokinetic simulations display a high degree of morphological symmetry, to which the PIC simulations converge when $\beta_i B_g / \delta B_0 \gtrsim 1$ and $B_g / \delta B_0 \gg 1$.  In the regime of convergence, the reconnection rate, relative energy conversion, and overall magnitudes are found to match well between the PIC and gyrokinetic simulations, implying that gyrokinetics is capable of making accurate predictions well outside its regime of formal applicability. These results imply that in the large guide field limit many quantities resulting from the nonlinear evolution of reconnection scale linearly with the guide field. 

\end{abstract}


\maketitle 

\emph{Introduction}---Magnetic reconnection is a ubiquitous phenomenon that likely plays a significant role in the transport of energy in the solar corona, solar wind, geospace environment, and magnetic confinement fusion devices such as ITER \cite{Zweibel:2009,Yamada:2010,Ji:2011}. Many of these environments are in the strongly magnetized, weak shear limit, where particle-in-cell (PIC) simulations become challenging.

The gyrokinetic (GK) system of equations has a long history in the magnetic confinement fusion community, e.g., \cite{Rutherford:1968,Taylor:1968,Antonsen:1980,Frieman:1982}, and has been recently applied to space and astrophysical plasmas, e.g., \cite{Howes:2006,Howes:2011b,TenBarge:2013a}. Fundamentally, GKs assumes linear frequencies are small compared to the cyclotron frequency and that the plasma is strongly magnetized. Applying this ordering to the Vlasov-Maxwell system results in the GK equations. 

Gyrokinetics has been used to model magnetic reconnection directly, e.g., \cite{Rogers:2007,Perona:2010,Numata:2010,Numata:2011,Pueschel:2011,Zocco:2011,Perona:2012,Zacharias:2012,Loureiro:2013}, and indirectly in simulations of magnetic confinement plasmas; however, the limitations of GKs regarding reconnection have not been thoroughly explored. Specifically, reconnection events often involve important physics that may violate the GK orderings, such as large amplitude, high frequency fluctuations and instabilities. 

In this letter, we perform the first detailed comparison between 2.5D magnetic reconnection as simulated by a GK code, AstroGK, and fully electromagnetic PIC codes, NPIC and VPIC, to determine if the full kinetic, PIC, description converges to GKs in the limit of strong guide magnetic field. To address this question, we initialize as nearly as possible identical simulations in PIC and AstroGK and examine both a fusion and solar corona relevant $\beta_i = 8 \pi n_{0i} T_{0i} / B_g^2 = 0.01$ and solar wind relevant $\beta_i = 1$. In both cases, the ratio of the guide to equilibrium reconnection magnetic field is scanned in PIC and comparisons are made between the simulation results. Reconnection rates, relative energy transport, and overall magnitudes are found to be well predicted by GKs, even in the weaker guide field cases examined. The morphology of the PIC reconnection results are found to converge to GKs in the limit of strong guide field, but the convergence has a linear $\beta_i$ dependence. The correct prediction of overall magnitudes by GKs implies that the reconnection results scale linearly with guide field strength. This is an important result since it implies that a single simulation can represent a range of large guide fields when properly scaled.

\emph{Gyrokinetics}---The GK system of equations is based upon a multi-scale expansion in which small-scale fluctuations vary on the linear ($\omega$) time-scale while macroscopic (background) quantities vary on the slow transport time-scale \cite{Rutherford:1968,Taylor:1968,Antonsen:1980,Frieman:1982,Howes:2006}. The basic assumptions of this system are: weak coupling, low frequency compared to the cyclotron frequency,  strong magnetization, and small fluctuations. We define the fundamental ordering parameter of GK to be $\epsilon = \omega / \Omega_{ci} \ll 1$ and all fluctuating quantities obey this ordering, e.g., $\delta F / F_0 \sim \delta B / B_g \sim \delta U_s / v_{ts} \sim \epsilon$, where $\delta$ quantities represent the fluctuating portion, $B_g$ is the equilibrium, guide magnetic field, $\Omega_{cs} = q_s B_g / m_s c$ is the gyrofrequency, and $v_{ts} = \sqrt{2 T_{0s} / m_s}$ is the thermal speed. When applied to the Vlasov-Maxwell system of equations, this ordering results in the GK equations. The ordering averages over the fast cyclotron motion of the particles and thus removes all cyclotron physics and the fast magnetosonic wave, reduces the phase space dimensionality from six to five, and describes the evolution of rings rather than particles. The ordering also enforces charge neutrality and magnetic moment conservation in the absence of collisions. Note that this ordering retains finite Larmor radius (FLR) effects, non-linear physics, and collisionless wave-particle damping. The details of this system of fully non-linear equations can be found in \cite{Frieman:1982,Howes:2006}. As noted above, background quantities such as $F_0$ are assumed to vary on the slow, transport time-scale, which is one order higher in $\epsilon$ than typically retained in the GK system. Therefore, this GK formulation represents a $\delta F$ system, meaning the background distribution of particles is not evolved. 

\emph{Code Descriptions}---The Astrophysical Gyrokinetics code (AstroGK) \cite{Numata:2010} has been chosen to represent the GK aspect of this study. AstroGK is the slab geometry sibling of the toroidal geometry code GS2 \cite{Kotschenreuther:1995,Dorland:2000}, which has been used extensively by the fusion community. AstroGK is an Eulerian continuum code with triply periodic boundary conditions that solves the five-dimensional, electromagnetic gyroaveraged Vlasov-Maxwell equations. The background distribution functions for both species are Maxwellian.  To obviate the numerical complexity of calculating asymptotically small values, all fluctuating quantities are normalized by $\epsilon$ to make them $\mathcal{O}(1)$, e.g., $\delta B^{AGK} = \delta B / B_g \epsilon$. Note that $\epsilon$ is neither a fixed nor user chosen parameter. Due to the GK ordering and the normalization employed in AstroGK, a value for $\epsilon$ must be chosen to make contact with reality and PIC. A natural choice is $\epsilon = \delta B_0 / B_g \equiv 1 / \hat{B}_g$, where $\delta B_0$ is the upstream magnitude of the reconnecting component of the magnetic field. For the purposes of this comparison, AstroGK is run in the fully collisionless limit. 

The first principles electromagnetic relativistic kinetic PIC codes NPIC \cite{Daughton:2003} and VPIC \cite{Bowers:2008a,Bowers:2009} have been chosen to represent the full, unordered kinetic system. NPIC uses an algorithm described in detail by \citet{Forslund:1985} and is a two spatial dimension Lagrangian particle code that integrates the full Maxwell-Boltzmann system of equations in 5D phase space and employs a semi-implicit scheme for improved efficiency and better noise properties. VPIC is a full, 6D Lagrangian particle code that employs state-of-the art computational techniques to maximize efficiency and scaling. No explicit collision operator is applied to either PIC code for the runs contained herein; however, the particle noise and coarse graining inherent to PIC simulations effectively smooths velocity space in a manner similar to collisions.

\emph{Simulation Details}---For this study, both systems are represented by 2.5D simulations of an electron-ion plasma in the $x$-$y$~plane, with a guide magnetic field in the out-of-plane direction, $\hat{\V{z}}$. A reduced mass ratio of $m_i / m_e = 25$ is used, and all simulations initialize $T_{0i} = T_{0e} = T_0$ and $n_{0i} = n_{0e} = n_0$. Both systems employ a fully periodic domain in the $x$-$y$~plane, requiring that two current sheets be present in all simulations. We here focus on one of the current sheets. For $\beta_i = 0.01~(1)$, both simulations use simulation domains $L_x=L_y = 40 \pi~(20 \pi) \rho_i$ and initial current sheet half widths $w = 2~(1) \rho_i$, where $\rho_i = v_{ti} / \Omega_{ci}$ is the ion gyroradius. The parameters employed for the PIC runs are as follows for $\beta_i = 0.01~(1)$: $\hat{B}_g = 5, 10, 20, 50~(1, 5, 10)$; $n_x = n_y = n = 1024, 1024, 2048, 2048~(1280)$; $n_{ppc} = 1000~(2000, 4000, 4000)$; $\omega_{pe} / \Omega_{ce} = 0.8~(4)$, where $n_{ppc}$ is the number of super particles per cell and $\omega_{pe} = \sqrt{4 \pi n_{0e} q_e^2 / m_e}$. The AstroGK simulations employ $n = 1024$ and $1280$ with velocity space grids $(n_\lambda, n_E) = (32, 16)$ and $(16,8)$ for $\beta_i = 0.01$ and $1$ respectively, where $n_\lambda$ and $n_E$ are pitch angle and energy grids. To keep $\beta_i$ fixed, the PIC simulations fix the magnitude of $B_g$ and reduce $\delta B_0$ to achieve the effect of increasing guide field strength. This approach necessarily implies that the signal-to-noise ratio (SNR) becomes increasingly poor as $\hat{B}_g$ is increased. Note that $\beta_i < m_i / m_e$ for $\beta_i = 0.01$, implying that no dispersive wave modes are present, and the reconnection rate has been predicted to be slow in this regime \cite{Rogers:2001}.

The initial magnetic geometry employed for this comparison is a force-free equilibrium of the form
\begin{equation}
\V{B} = \delta B_0 \tanh{(y/w)} \hat{\V{y}}+ \sqrt{B_g^2 + \delta B_0^2 \cosh^{-2}{(y/w)}} \hat{\V{z}}.
\end{equation}
The ions are stationary and the electrons have a net drift velocity $\delta U_e$ consistent with Ampere's law. This system has recently been studied in detail by \citet{Liu:2013,Liu:2014}.


A $1\%$ GEM-like \cite{Birn:2001} perturbation to $B_x$ is added to the force-free equilibrium to initiate reconnection. Since PIC simulations inherently suffer from particle noise, the first twenty non-zero Fourier modes in $\hat{\V{x}}$ and $\hat{\V{y}}$ of the AstroGK runs have been populated with uniform random noise whose spectra and amplitude approximate the initial noise present in the parallel vector potential of the PIC simulations---approximately $0.2\%$ of the equilibrium.

\begin{figure}[top]
\includegraphics[width=\linewidth]{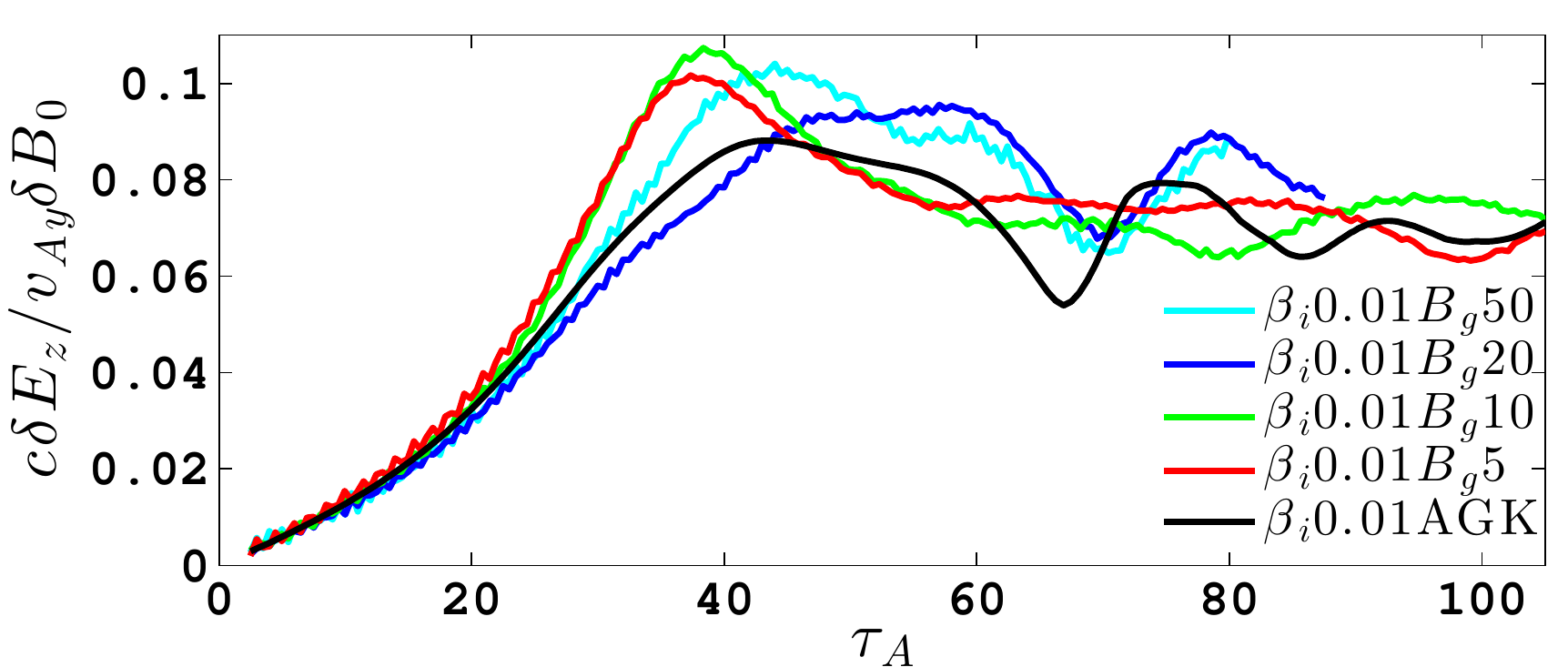}
\includegraphics[width=\linewidth]{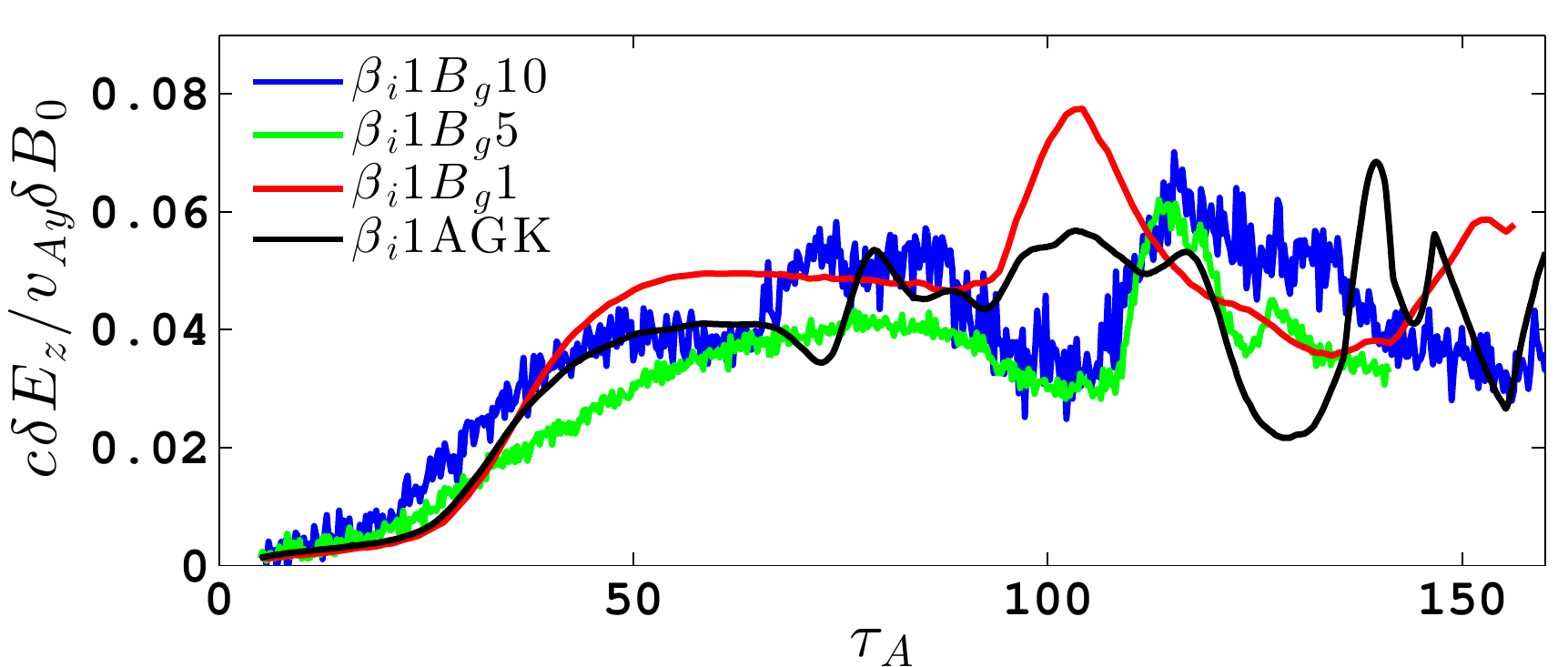}
\caption{Reconnection rate versus time for the, a), $\beta_i = 0.01$ and, b), $\beta_i = 1$ simulations. Black represents AstroGK and red, green, blue, and cyan represent increasing values of $\hat{B}_g$ for the PIC simulations.}\label{fig:recon_rate}
\end{figure}

\emph{Reconnection Rate}---We begin our analysis by examining the reconnection rates of the runs, which are presented in Figure~\ref{fig:recon_rate}. Time is normalized to the in-plane Alfv\'{e}n crossing time, $\tau_A = t v_{Ay} / w$, where $v_{Ay} = \delta B_0 / \sqrt{4 \pi n_0 m_i}$. The reconnection rate is fast and agrees well for all runs, but the rate does show a $\beta_i$ dependence. As noted in \cite{Liu:2014}, reconnection remains fast in fully kinetic systems, even in the regime of $\beta_i < m_i / m_e$, where slow reconnection is observed in two fluid models. To confirm that the observed reconnection rate is indeed fast for the $\beta_i = 0.01$ case, a scaling study was performed using AstroGK wherein the box size was halved and doubled relative to the simulation presented herein while all other parameters were held fixed. The peak reconnection rates were found to be $0.10$ and $0.12$ for the half and double size domains, which is comparable to the observed rate of $0.088$ for the nominal run. A more detailed study spanning a wider range of simulation domains and plasma parameters will be presented elsewhere. The sharp gradients and differences in reconnection rate apparent for the $\beta_i = 1$ case for $\tau_A \gtrsim 80$ are due to the current sheets thinning and elongating to the point of being unstable to the formation of secondary islands, whose development is highly sensitive to noise in the simulation. Thus, moderate disagreement in the reconnection rate at these late times is not unexpected.

\begin{figure}[top]
\includegraphics[width=\linewidth]{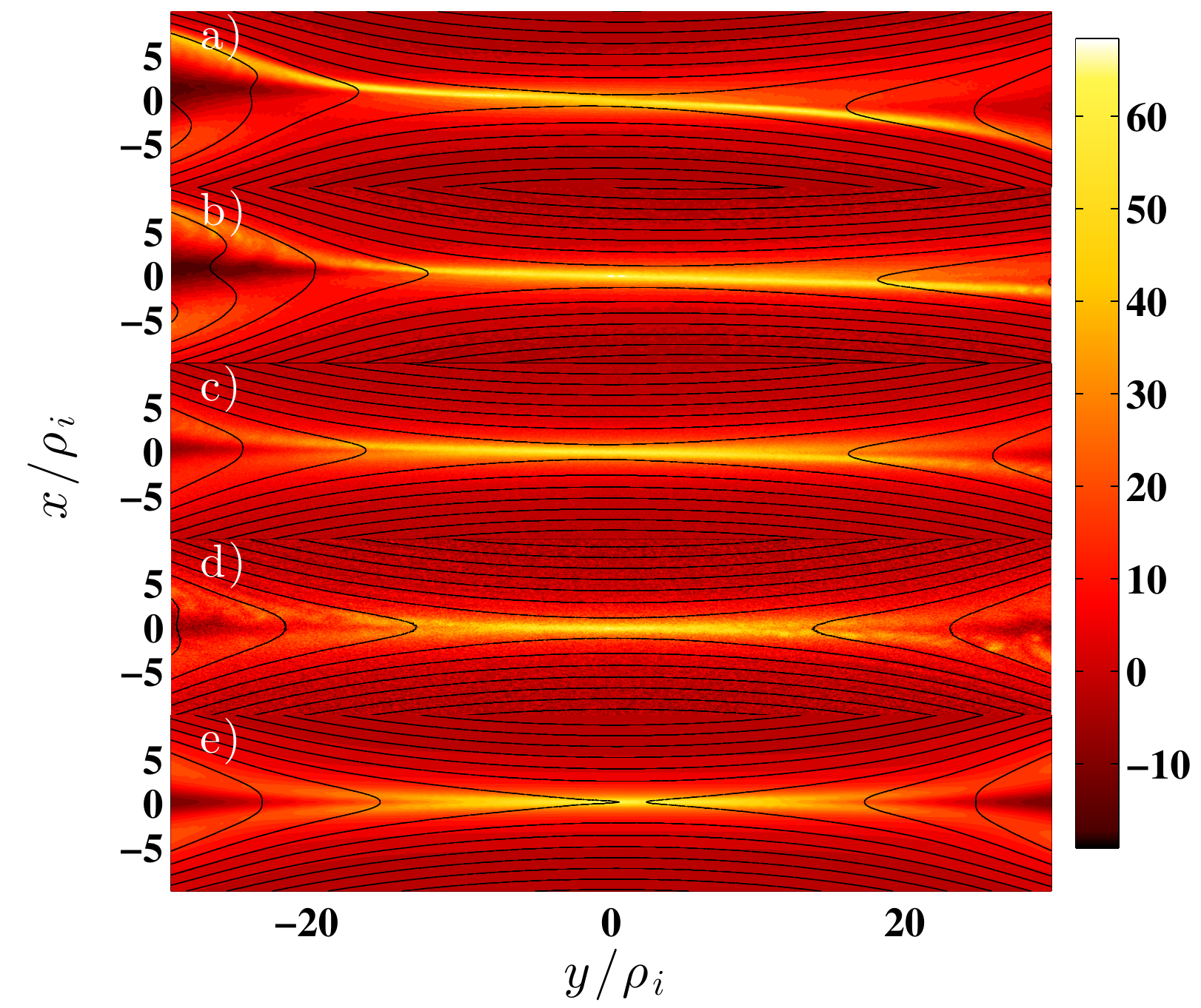}
\caption{Contours of $J_z$ normalized to AstroGK units at $\tau_A = 80$ for $\beta_i = 0.01$. a)-d) are PIC simulations with $\hat{B}_g = 5,10,20,$ and $50$ respectively and e) is AstroGK.}\label{fig:contour1}
\end{figure}

\begin{figure}[top]
\includegraphics[width=\linewidth]{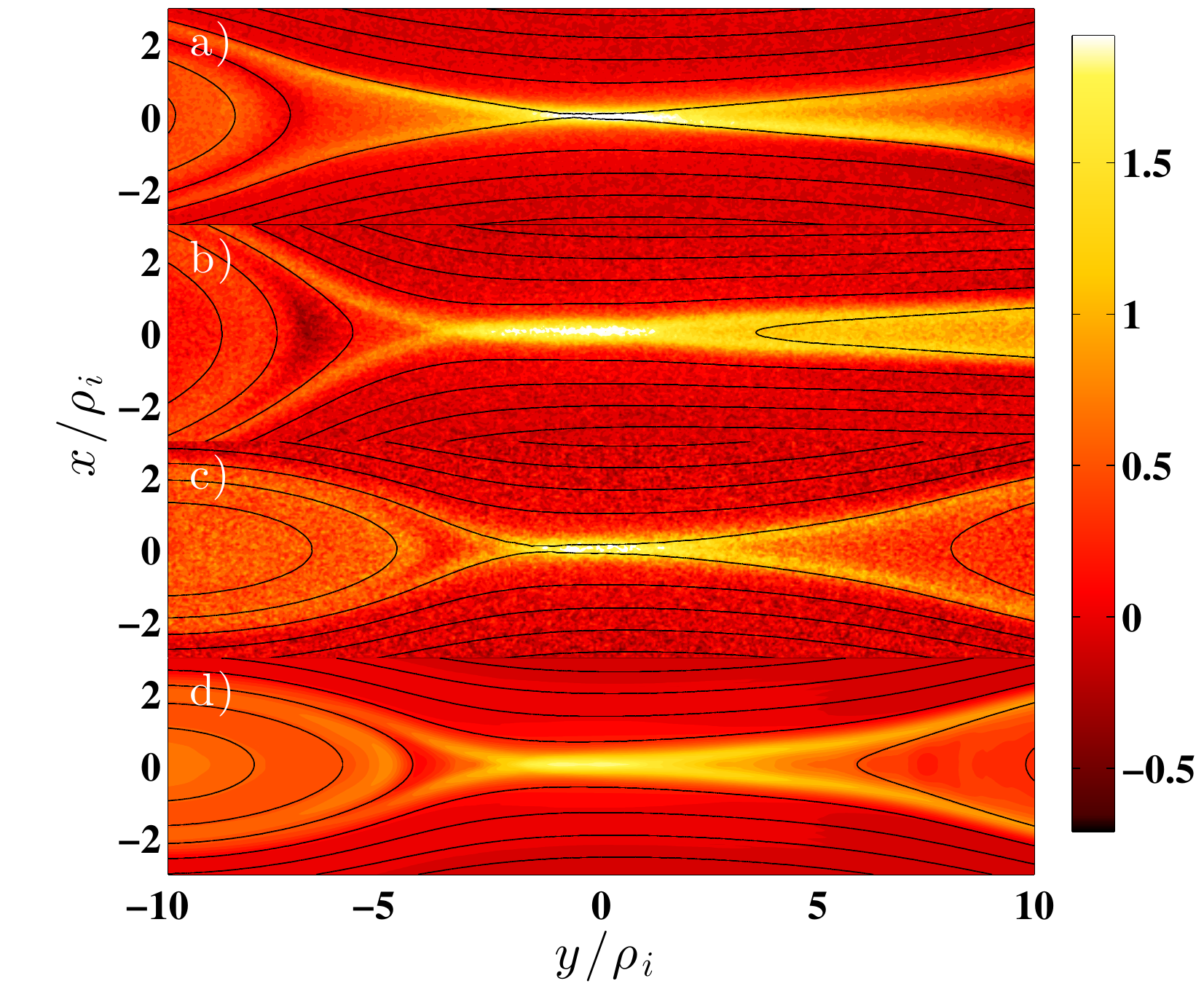}
\caption{Contours of $J_z$ normalized to AstroGK units at times prior to secondary island formation and centered on the x point for $\beta_i = 1$. a)-c) are PIC simulations with $\hat{B}_g = 1,5,$ and $10$ respectively and d) is AstroGK.}\label{fig:contour2}
\end{figure}

\emph{Morphology}---For the purpose of making detailed comparisons of the structure and magnitude of relevant physical quantities, we will focus on times $\tau_A = 80$ for $\beta_i = 0.01$ and the times prior to secondary island formation for $\beta_i = 1$ respectively---secondary islands form at $\tau_A \simeq 88, 90, 64,$ and $73$ for $\hat{B}_g = 1,5,10$ and AstroGK respectively. These times have been chosen because reconnection is well developed and the morphology is unaffected by secondary islands. Figures~\ref{fig:contour1} and \ref{fig:contour2} present the out of plane current, $J_z$, from PIC and AstroGK at the indicated times. To decrease the significance of the noise in the high $\hat{B}_g$, $\beta_i = 1$ runs, the current has been averaged over $ \Delta \tau_A = 0.5$ and $1$ for $\hat{B}_g = 5$ and $10$ respectively. These figures were produced by converting the PIC results to AstroGK units, which linearly normalizes the PIC results by the strength of $\hat{B}_g$. Immediately clear from Figure~\ref{fig:contour1} is that the morphology converges with increasing $\hat{B}_g$ toward the highly symmetric AstroGK result for $\beta_i = 0.01$. In contrast, the $\beta_i = 1$ PIC simulations presented in Figure~\ref{fig:contour2} display a high degree of morphological convergence even at $\hat{B}_g = 1$, although some asymmetry in the current is visible for $\hat{B}_g = 1$.

The physical explanation for the morphological differences follows from pressure balance across the current sheet, $\delta P^{tot} + B_g \delta B_z / 4 \pi + \delta B^2 / 8 \pi = const$. $\delta P^{tot}$ has been observed to maintain quadrupolar (odd) symmetry across reconnection layers for nonzero $B_g$, e.g., \cite{Kleva:1995,Rogers:2003}, while the final term on the LHS is manifestly even.  Therefore, the balance of these two terms determines the symmetry of $\delta B_z$ \cite{Rogers:2003}, and the density is anti-correlated to $\delta B_z$ along the separatrices, which determines the structure of the current layer in the same region. In the limit $B_g \gg \delta B_0$, it is reasonable to assume the ordering $\delta P / P_0 \sim \delta B_0 / B_g$. Therefore, 
\begin{equation}\label{eq:pbal}
\frac{\delta B_z^{odd}}{\delta B_z^{even}} \sim \beta_i \frac{B_g}{\delta B_0}.
\end{equation}
Thus, the $\beta_i = 0.01$ set of PIC simulations exhibit significant but decreasing asymmetry up to $\hat{B}_g \sim 50$, where equation~\eqref{eq:pbal} is $\mathcal{O}(1)$, while the $\beta_i = 1$ simulations exhibit symmetry for $\hat{B}_g > 1$, and the equations simulated in AstroGK order equation~\eqref{eq:pbal} formally infinite.

\emph{Magnitude}---As can be seen in Figures~\ref{fig:contour1} and \ref{fig:contour2}, the magnitude of the normalized $J_z$ agrees very well with the results from AstroGK. This agreement in magnitude extends across all quantities, including magnetic field strengths, outflow velocities, temperatures, and anisotropies. Table~\ref{tab:rms} summarizes this finding by examining the ratio of the AstroGK normalized PIC quantities to the same AstroGK quantity. The root mean square (RMS) value of each quantity is found over the same regions presented in Figures~\ref{fig:contour1} and \ref{fig:contour2}. The temperature anisotropy, $T_\perp / T_\parallel$, is computed with respect to the total magnetic field. To construct this quantity in GKs, the background temperature and guide field are assumed to be $1 / \epsilon$. For low values of $\hat{B}_g$, this construction can lead to cases wherein $T_\parallel \rightarrow 0$ at some locations in the simulation domain. This phenomenon is responsible for the significant disagreement between AstroGK and PIC for run $\beta_i0.01Bg5$.  All ratios in Table~\ref{tab:rms} agree relatively well, aside from the higher $\hat{B}_g$ runs, whose values are dominated by small-scale noise fluctuations. AstroGK does consistently underestimate some quantities such as density for $\beta_i = 1$ and $v_z^i$; the reason for this disagreement will be explored in more detail in a forthcoming paper. There are two significant implications stemming from the overall good agreement: 1) The linear normalization of the PIC results by the strength of $\hat{B}_g$ implies that quantities associated with reconnection scale linearly with $\hat{B}_g$ in the $\hat{B}_g \gg 1$ limit, a heretofore unobserved result.  2) AstroGK accurately predicts the magnitudes of many quantities well outside its regime of formal applicability. Agreement with the PIC results at low $\hat{B}_g$ and $\beta_i$ implies that AstroGK accurately predicts such phenomena as suprathermal outflow velocities and $\mathcal{O}(1)$ density and magnetic fluctuations, predictions that violate the ordering assumptions of GKs. 

\begin{table}[t]
\caption{Ratios of the RMS of each value from PIC to the RMS AstroGK value over the regions presented in Figures~\ref{fig:contour1} and \ref{fig:contour2}.}\label{tab:rms}
\begin{center}
\begin{tabular}{|c||c|c|c|c|c|c|c|}
\hline \hline
        Run & $J_z$  & $v_z^i$ & $v_y^e$ & $B_x$ & $n_e$ & $(T_\perp / T_\parallel)^i$ & $(T_\perp / T_\parallel)^e$\\
\hline
$\beta_i0.01B_g5$ & $1.1$  &  $1.2$ & $0.90$ & $1.0$ & $0.92$& $0.97$ & $0.012$\\
\hline
$\beta_i0.01B_g10$ & $1.1$ &  $1.2$ & $0.94$ & $0.93$ & $0.98$& $1.0$ &$0.95$\\
\hline
$\beta_i0.01B_g20$ & $1.0$   &  $1.2$ & $1.0$ & $0.85$& $1.1$& $1.0$ & $0.99$\\
\hline
$\beta_i0.01B_g50$  & $1.0$   &  $1.5$ & $2.4$ & $0.95$& $2.0$& $1.0$ & $1.0$\\ 
\hline
$\beta_i1B_g1$  & $1.1$   & $1.4$ & $1.1$ & $1.0$& $1.3$& $0.84$ & $0.92$\\ 
\hline
$\beta_i1B_g5$  & $1.1$    & $1.4$ & $1.1$ & $1.2$& $1.3$& $0.99$ & $1.0$\\ 
\hline
$\beta_i1B_g10$  & $1.1$  & $0.98$ & $1.1$ & $0.99$& $1.8$& $1.0$ & $1.1$\\ 
\hline
\end{tabular}
\end{center}
\end{table}

\emph{Energy}---For the PIC simulations, $E = \int \, d^3 \mathbf{r} / V \left(B^2 / 8 \pi + \sum_s m_s n_{0s} \delta U_s^2 / 2 + 3 P_s / 2\right)$ is a well conserved quantity, where all bulk velocity, $U$, is assumed to be fluctuating and the small contribution from the electric field has been neglected. However, the conservation of this expression for the energy assumes a collisional system, which the effective collisionality inherent to PIC simulations provides. Since AstroGK is a continuum code and being run in the collisionless limit for this comparison, the above form of the energy is not a well conserved quantity. This is because energy in velocity space cascades toward sharper gradients through linear and non-linear phase space mixing, which implies that energy is transferred to progressively higher moments as the system evolves unless collisions smooth velocity space. The relevant conserved quantity in GKs is the generalized energy \cite{Howes:2006,Schekochihin:2009}
\begin{equation}
\begin{split}
&W = \int \,\frac{d^3\mathbf{r}}{V} \left[\frac{\delta B^2}{8 \pi}  + \sum_s \int \,d^3 \mathbf{v} \frac{T_{0s} \delta f_s^2}{2 F_{0s}}\right] =\\
& \int \,\frac{d^3\mathbf{r}}{V} \left[\frac{\delta B^2}{8 \pi}  + \sum_s \frac{1}{2} m_s n_{0s} \delta U_s^2 + \int \,d^3 \mathbf{v} \frac{T_{0s} \delta \tilde{f}_s^2}{2 F_{0s}}\right],
\end{split}
\end{equation}
where $\delta \tilde{f}_s$ is the portion of the perturbed distribution that does not contribute to the bulk velocity. We will treat the $\delta \tilde{f}_s$ term as equivalent to the perturbed portion of the $3 P_s / 2$ term from PIC.

To avoid choosing a value of $\epsilon$ to compare the evolution of the energy in each system, we examine only the fluctuating energy in PIC by removing $B_g$ and the initial background $P_s$, $\delta E = \int \, d^3 \mathbf{r} / V \left(\delta B^2 / 8 \pi + \sum_s m_s n_{0s} \delta U_s^2 / 2 + 3 \delta P_s / 2\right)$. This quantity is well conserved for most of the PIC simulations but is conserved only to the $10-20\%$ level for all values of $\hat{B}_g$ at $\beta_i = 1$ and $\hat{B}_g = 50$ for $\beta_i = 0.01$. The poor conservation is due to poor SNR in these runs and appears primarily in the electron bulk kinetic and thermal energies.

\begin{figure}[top]
\includegraphics[width=\linewidth]{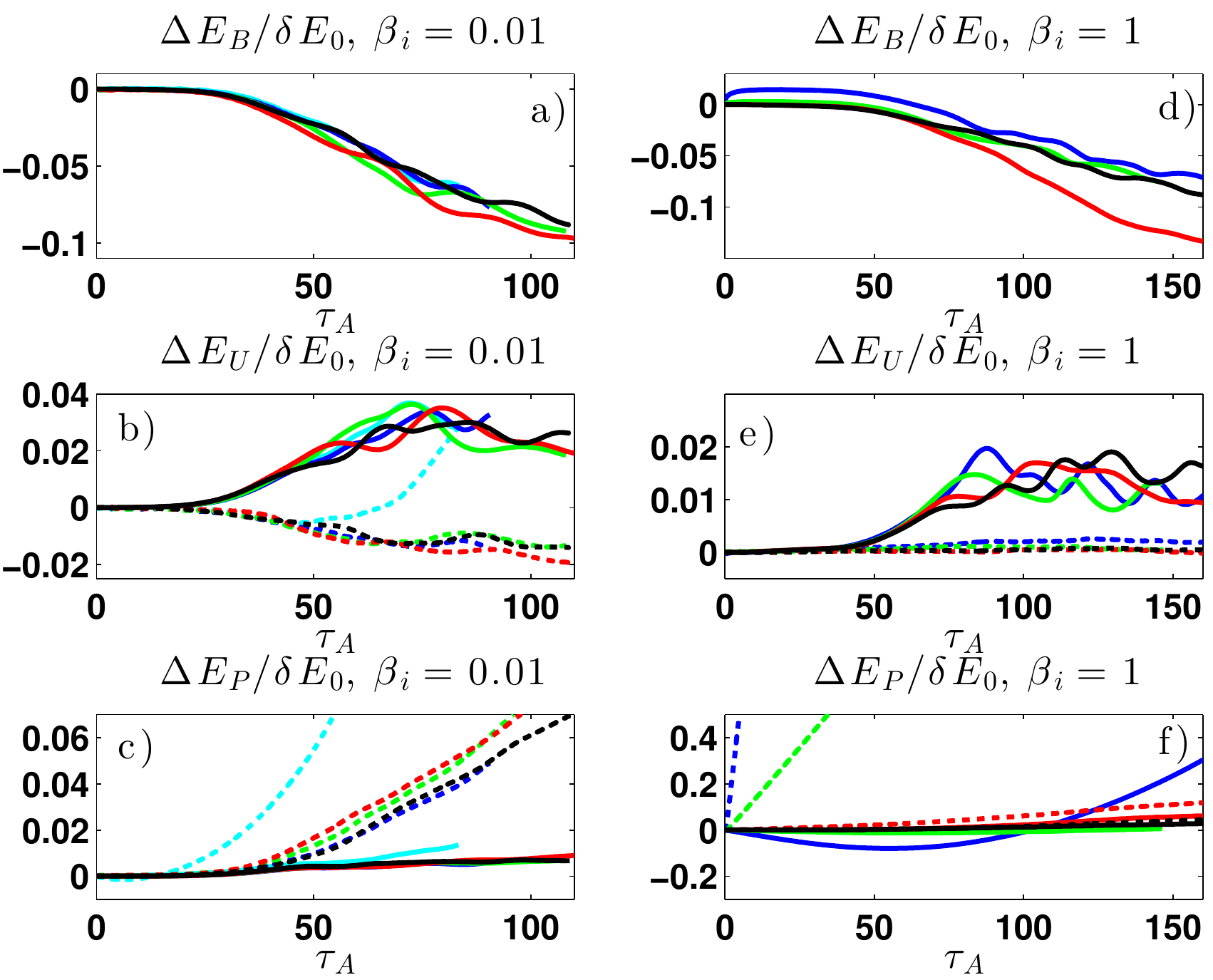}
\caption{Plots of the change in the, a) and d), magnetic, b) and e), bulk kinetic and, c) and f), thermal energies from their initial values for, a)-c), $\beta_i = 0.01$ and, d)-f), $\beta_i = 1$. Solid lines indicate ion quantities and dashed are electron. All energies are normalized to the total initial energy in the system. Color scheme is the same as Figure~\ref{fig:recon_rate}. }\label{fig:energy}
\end{figure}

The evolution of the fluctuating magnetic, bulk kinetic, and thermal energies from their respective initial energies are plotted in Figure~\ref{fig:energy} for all of the simulations. All energies are normalized to the total fluctuating energy at the beginning of each simulation, $\delta E_0$. Aside from the  high $\hat{B}_g$ run having poor SNR, the evolution of the magnetic, bulk kinetic, and ion thermal energies show excellent agreement across all simulations for which $\hat{B}_g > 1$. The significant disagreement for the electron thermal energy for $\beta_i = 1$ is dominated by numerical heating of the entire simulation domain. Note that although the relative energy change matches well across all runs, the free energy in each simulation scales with $ \hat{B}_g^{-2}$. Therefore, the large $\hat{B}_g$ cases have significantly less energy available to accelerate and heat particles.

\emph{Conclusions}---In the large guide field limit, we have shown that the fully kinetic particle-in-cell simulations converge to gyrokinetic results for 2.5D magnetic reconnection. Morphological convergence was shown to require $\beta_i B_g / \delta B_0 \gtrsim 1$ and $B_g / \delta B_0 \gg 1$, implying much stronger guide fields are needed at low $\beta_i$ to achieve convergence, where $\delta B_0$ is the reconnecting field. Reconnection rates, relative energy conversion from magnetic to bulk kinetic and thermal energies, and appropriately normalized overall magnitudes were found to match well for all values of $\beta_i$ and $B_g / \delta B_0 > 1$. The observed amplitude scaling implies that gyrokinetics is capable of making accurate predictions well outside its formal regime of applicability, and that magnetic reconnection in the large guide field limit produces quantities that scale linearly with the guide field strength, which implies that a single simulation can be scaled to represent a range of guide fields. This study validates the use of gyrokinetics as alternative means of studying reconnection in the strongly magnetized kinetic regime. Future work will examine these simulations in greater detail, including their temporal evolution, comparison to linear theory, and development of secondary instabilities.


Acknowledgments---The authors thank James Drake for helpful discussions. This work was supported by the US DOE grant DEFG0293ER54197, NSF CAREER AGS-1054061, NASA grant NNH11CC65C, and NASA's Heliophysics Theory Program. This work used the Extreme Science and Engineering Discovery Environment (XSEDE), which is supported by NSF grant numbers PHY090084 and OCI-1053575, with runs performed on Kraken at the National Institute for Computational Sciences, Stampede at the Texas Advanced Computing Center. Additional simulations were performed on Pleiades provided by NASA's HEC Program and with Los Alamos Institutional Computing resources.

%

\end{document}